\documentclass[12pt]{article}
\usepackage{latexsym}
\newtheorem{theorem}{Theorem}

\date{}

\begin{document}

\begin{center}
{\Large \bf  Partner symmetries and non-invariant solutions of
four-dimensional heavenly equations}\\[4mm]
{\large \bf A A Malykh$^1$, Y Nutku$^3$ and
M B Sheftel$^{1,2,3}$}
\end{center}
\vspace{1mm}
{\flushleft $^1$ Department of Higher Mathematics,
North Western State Technical University, Millionnaya St. 5,
191186, St. Petersburg, Russia
\\[1mm]
$^2$ Department of Physics, Bo\u{g}azi\c{c}i University, 34342
Bebek, Istanbul, Turkey
\\[1mm] $^3$ Feza G\"{u}rsey Institute, PO Box 6, \c{C}engelk\"{o}y,
81220 Istanbul, Turkey
\vspace{1mm}
\\ E-mail: specarm@online.ru,
nutku@gursey.gov.tr, sheftel@gursey.gov.tr and
mikhail.sheftel@boun.edu.tr
\vspace{3mm}
\\{\bf Abstract}
\\ We extend our method of partner symmetries to the hyperbolic
complex Monge-Amp\`ere equation and the second heavenly equation
of Pleba\~nski. We show the existence of partner symmetries and
derive the relations between them. For certain simple choices of
partner symmetries the resulting differential constraints together
with the original heavenly equations are transformed to systems of
linear equations by an appropriate Legendre transformation. The
solutions of these linear equations are generically non-invariant.
As a consequence we obtain explicitly new classes of heavenly
metrics without Killing vectors. \vspace{3mm}
 }

\section{Introduction}
\setcounter{equation}{0}

After advances in Twistor Theory it became natural to consider
Ricci-flat metrics on $4$-dimensional complex manifolds. In his
pioneering paper \cite{pleb} Pleba\~nski introduced his first and
second heavenly equations for a single potential governing such
metrics. From solutions of these equations we obtain the
corresponding heavenly metrics which give solutions of the complex
vacuum Einstein equations possessing the property of
(anti-)self-duality. The problem here is to reduce these solutions
to $4$-dimensional real metrics with Lorentzian signature. There
are two important real cross sections of the complex metrics
governed by the first heavenly equation, namely K\"ahler metrics
with Euclidean or ultra-hyperbolic signature. The first heavenly
equation in these cases coincides with the elliptic and hyperbolic
complex Monge-Amp\`ere equation ($CMA$) respectively which have
applications to important problems in physics and geometry. In
particular, for the elliptic case of $CMA$ some solutions can be
interpreted as gravitational instantons. The most important
gravitational instanton is the Kummer surface $K3$ \cite{ahs}. The
explicit construction of $K3$ metric is a challenging problem. One
of the basic difficulties is that the metric should have no
Killing vectors and hence the corresponding solution of $CMA$
should have no symmetries, i.e. be a non-invariant solution.

This was our motivation to study the problem of constructing
non-in\- variant solutions of complex Monge-Amp\`ere equations. In
the elliptic case we have recently developed the method of partner
symmetries appropriate for this problem and obtained certain
classes of non-invariant solutions of $CMA$ and corresponding
heavenly metrics with no Killing vectors \cite{mns,mnsgr}. The
starting point of the method was the observation that the
determining equation for symmetries of the elliptic $CMA$ can be
presented in the form of a total divergence. This allowed us to
introduce locally a potential variable, and another key
observation was that the potential again satisfied the same
determining equation, {\it i.e.} also was a symmetry. We called
such a pair of original symmetry and its potential {\it partner
symmetries}. The equations relating partner symmetries are
therefore recursion relations mapping any symmetry again into a
symmetry of $CMA$ but the corresponding recursion operator is
non-local. Thus if we apply it to a local symmetry then a
non-local symmetry will be generated. To avoid the explicit use of
non-local symmetries, we consider both partner symmetries as local
symmetries and the relation between them as invariance condition
with respect to a certain resulting non-local symmetry which we
never explicitly put into play. Hence the resulting solutions,
though invariant with respect to some non-local symmetry, will be
non-invariant solutions in the usual sense and hence generate
metrics without Killing vectors. This is closely related to the
approach of Dunajski and Mason \cite{dm} though their `hidden'
symmetries belong to a more special class of non-local symmetries
than those constructed from the partner symmetries.

For any particular choice of partner symmetries the relations
between them become differential constraints compatible with the
original $CMA$ equation. We discovered certain useful choices of
partner symmetries when $CMA$ together with the differential
constraints could be linearized by a Legendre transformation.
Solving linear equations and using their solutions in the Legendre
transform of the metric we obtained explicitly some classes of
Riemannian metrics with Euclidean signature and anti-self-dual
curvature that did not admit any Killing vectors. It is worth
noting that linearization of particular solution manifolds of PDEs
by the Legendre transformation was suggested in \cite{ivroc}.

In the present paper we extend our method of partner symmetries to
two more $4$-dimensional heavenly equations: hyperbolic complex
Monge-Amp\`ere equation ($HCMA$), which is the other interesting
real cross section of the first heavenly equation, and the second
heavenly equation of Pleba\~nski. Our method of partner symmetries
works because determining equations for symmetries of all these
heavenly equations have the structure of total divergence, and the
potential for a symmetry is itself a symmetry. Therefore we again
have pairs of partner symmetries for all these heavenly equations.
We find such choices of partner symmetries when the original
heavenly equation together with differential constraints arising
from relations between partner symmetries can be linearized by the
Legendre transformation. We present solutions of linear equations
and the corresponding metrics without Killing vectors.

We study the hyperbolic complex Monge-Amp\`ere equation in the
first part of the paper and the second heavenly equation in the
second part.

In section \ref{sec-hyperbol}, for the sake of completeness, we
show that if the K\"ahler potential satisfies $HCMA$ the
corresponding K\"ahler metric has ultra-hyperbolic signature. In
section \ref{sec-partner-hcma} we derive the relations defining
partner symmetries from divergence structure of the determining
equation for symmetries of $HCMA$ and make our choices of partner
symmetries. In section \ref{sec-legendr-hcma} we consider partial
Legendre transformation of $HCMA$ and the relations between
partner symmetries in the case when both partner symmetries
coincide. We also obtain the Legendre transform of the K\"ahler
metric. Making further choices of partner symmetries as
translational or dilatational symmetry we arrive at systems of
linear PDEs for which we obtain explicitly non-invariant
solutions. Using them in the Legendre transformed metric we obtain
explicit 4-dimensional metrics without Killing vectors which is
justified in section \ref{sec-killing}.

In section \ref{sec-sym-heaven} we consider the second heavenly
equation with the corresponding heavenly metrics. In subsection
\ref{sec-partner-heaven} we show the existence of partner
symmetries and derive the relations defining them from divergence
structure of the determining equation for symmetries. In
subsection \ref{sec-pointsym} we present explicitly basis
generators of the total Lie algebra of point symmetries of the
second heavenly equation and the table of its commutation
relations. The Legendre transform of the second heavenly metric is
given in subsection \ref{sec-legendr-heaven}. In section
\ref{sec-translsym} we consider two simple choices of partner
symmetries. In subsection \ref{sec-equalsym} we discuss the
subcase when both of them coincide with the same translational
symmetry. The Legendre transformation of the heavenly equation and
differential constraints arising from this choice convert them to
a system of linear equations which is easily solved. In subsection
\ref{sec-highsym} we choose one of the partner symmetries to be
translational symmetry and the characteristic of the other one is
set equal to zero. The same Legendre transformation gives us again
a linear system which is easily solved. In the generic case all
these solutions are non-invariant and the corresponding metrics
have no Killing vectors.

In section \ref{sec-killing} we analyze the Killing equations for
both the K\"ahler and the second heavenly metrics. We find a first
integral of the Killing equations which is a first-order PDE for
the metric potential that contains all the information in
Killing's equations. We show that for our solutions this existence
condition for the Killing vector cannot be satisfied as it implies
functional dependence between independent variables.

\section{Hyperbolic complex Monge-Amp\`ere\\ equation and
ultra-hyperbolic metrics} \setcounter{equation}{0}
\label{sec-hyperbol}

In the famous paper of Pleba\~nski \cite{pleb} the Einstein vacuum
equations in the complex four-dimensional Riemannian space
together with the constraint of (anti-)self-duality are reduced to
the general complex Monge-Amp\`ere equation ($CMA$)
\begin{equation}
\left|
\begin{array}{cc}
\Omega_{pr} & \Omega_{ps}
\\ \Omega_{qr} & \Omega_{qs}
\end{array}
\right| = 1
\label{gcma}
\end{equation}
governing the metric
\begin{equation}
ds^2 = \Omega_{pr} dp dr + \Omega_{ps} dp ds + \Omega_{qr} dq dr +
\Omega_{qs} dq ds
\label{cmetr}
\end{equation}
where the `key function' $\Omega$ is a complex-valued function of
the complex variables $p,q,r,s$ and we skip the overall constant
factor $2$.

To restrict this general result to physically interesting cases,
we require that $\Omega = u$ where $u$ is a real-valued function
and the independent variables form two pairs of complex conjugate
variables: $p=z^1,\; q=z^2$, $r=\varepsilon\bar z^1$, $s=\bar z^2$
with $\varepsilon = \pm 1$. Then the field equation (\ref{gcma})
takes the form
\begin{equation}
u_{1\bar 1} u_{2\bar 2} - u_{1\bar 2} u_{2\bar 1} = \varepsilon
\label{ecma}
\end{equation}
and the metric (\ref{cmetr}) becomes
\begin{equation}
ds^2 = u_{1\bar 1} dz^1 d\bar z^1 + u_{1\bar 2} dz^1 d\bar z^2 +
u_{2\bar 1} dz^2 d\bar z^1 + u_{2\bar 2} dz^2 d\bar z^2
\label{metr}
\end{equation}
where the subscripts $i, \bar j$ denote partial derivatives with
respect to $z^i, \bar z^j$ respectively.

To determine the signature of the resulting metric (\ref{metr})
and its dependence on $\varepsilon$, we use the tetrad of the
Newman-Penrose co-frame $\{l,\bar l,m,\bar m\}$ \cite{gol,an}
corresponding to the metric (\ref{metr})
\begin{equation}
l = \frac{1}{\sqrt{u_{1\bar 1}}}\big(u_{\bar 11}dz^1 + u_{\bar 12}
dz^2 \big), \qquad m = \frac{1}{\sqrt{u_{1\bar 1}}}\, dz^2
\label{tetrad}
\end{equation}
with $\bar l$ and $\bar m$ being complex conjugate to $l$ and $m$.
Then using the field equation (\ref{ecma}) in the form
\[ \frac{u_{1\bar 2}u_{\bar 12}}{u_{1\bar 1}} = u_{2\bar 2}
- \frac{\varepsilon}{u_{1\bar1}} \]
we check that the metric
\begin{equation}
ds^2 = l \otimes \bar l + \varepsilon m \otimes \bar m
\label{invmetr}
\end{equation}
coincides with (\ref{metr}). The form (\ref{invmetr}) implies that
the metric (\ref{metr}) is Euclidean with the signature $(++++)$
for $\varepsilon = 1$ and it is ultra-hyperbolic with the
signature $(++--)$ for $\varepsilon = -1$. The difference comes
solely from the dependence of the field equation (\ref{ecma}) on
$\varepsilon$ whereas the metric has the same form (\ref{metr}) in
both cases.

In the first part of this paper we shall be interested in the
ultra-hyperbolic metric (\ref{metr}) governed by the hyperbolic
complex Monge-Amp\`ere equation ($HCMA$)
\begin{equation}
u_{1\bar 1} u_{2\bar 2} - u_{1\bar 2} u_{2\bar 1} = -1
\label{hcma}
\end{equation}
i.e. (\ref{ecma}) with $\varepsilon = -1$.

The relation between ultra-hyperbolic metrics and $HCMA$ was
demonstrated by a different method in \cite{lp}.

\section{Partner symmetries of the hyperbolic\\ complex Monge-Amp\`ere equation}
\setcounter{equation}{0} \label{sec-partner-hcma}

  The determining equation for symmetries of $HCMA$ is
the same as for the elliptic $CMA$ (\ref{ecma}) with $\varepsilon
= 1$ \cite{mns}
\begin{equation}
\Box(\varphi)=0,\qquad  \Box = u_{2\bar 2}D_1D_{\bar 1}+u_{1\bar
1}D_2D_{\bar 2} - u_{2\bar 1}D_1D_{\bar 2}-u_{1\bar 2}D_2D_{\bar
1}
\label{desym}
\end{equation}
where $\varphi$ denotes symmetry characteristic \cite{olv} and
$D_i, D_{\bar i}$ are operators of total derivatives with respect
to $z^i, \bar z^i$ respectively. Our starting point is the
divergence form of this equation
\begin{equation}
D_1L_2\varphi = D_2L_1\varphi
\label{div}
\end{equation}
where
\begin{equation}
L_1 = i (u_{1\bar 2}D_{\bar 1} - u_{1\bar 1}D_{\bar 2}), \quad L_2
= i (u_{2\bar 2}D_{\bar 1} - u_{2\bar 1}D_{\bar 2}) .
\label{L_i}
\end{equation}
Hence there locally exists the potential $\psi$ such that
\begin{equation}
\psi_1 = L_1\varphi = i (u_{1\bar 2}\varphi_{\bar 1} - u_{1\bar
1}\varphi_{\bar 2}),\quad \psi_2 = L_2\varphi = i (u_{2\bar
2}\varphi_{\bar 1} - u_{2\bar 1}\varphi_{\bar 2}).
\label{psi}
\end{equation}
Using this definition of $\psi$ and $HCMA$, a straightforward
check shows that $\Box(\psi)=0$ and thus $\psi$ is also a symmetry
characteristic of $HCMA$ together with $\varphi$. Thus equations
(\ref{psi}) are recursion relations for symmetries of $HCMA$.

We have called such a pair of mutually related symmetry
characteristics {\it partner symmetries} \cite{mns}. Formulas
(\ref{psi}) can be presented in the form
\begin{equation}
\psi=R\varphi=D_1^{-1}L_1\varphi,\qquad
\psi=R\varphi=D_2^{-1}L_2\varphi \label{R_op}
\end{equation}
where $R$ is recursion operator defined on the subspace of partner
symmetries.

Solving algebraically the equations complex conjugate to
(\ref{psi}) with respect to $\varphi_1, \varphi_2$ and using
(\ref{hcma}) we obtain
\begin{equation}
\varphi_1 = L_1\psi = i (u_{1\bar 2}\psi_{\bar 1} - u_{1\bar
1}\psi_{\bar 2}), \quad  \varphi_2 = L_2\psi = i (u_{2\bar
2}\psi_{\bar 1} - u_{2\bar 1}\psi_{\bar 2})
\label{phi}
\end{equation}
so that $\varphi$ is expressed through $\psi$ exactly in the same
way as $\psi$ is expressed through $\varphi$ in (\ref{psi}). This
is the basic difference with the elliptic $CMA$ for which we had
an extra minus sign: $\varphi_1 = -L_1\psi$, $\varphi_2 =
-L_2\psi$.

 The $HCMA$ itself emerges now as an algebraic consequence
of any three equations of the system (\ref{psi}), (\ref{phi}) on
the account of complex conjugate equations.

We will avoid operating explicitly with non-local symmetries and
use instead the equations (\ref{phi}) only with the point
symmetries $\varphi$ and $\psi$. Then these equations are
equivalent to an invariance condition for solutions of $HCMA$ with
respect to a non-local symmetry which is a linear combination of
$\psi$ and the non-local symmetry generated from $\varphi$ by the
recursion operator. Such solutions will still be non-invariant in
the usual sense.

We shall choose $\varphi$ and $\psi$ as characteristics of point
symmetries of $HCMA$. A general form of generators of point
symmetries of $HCMA$ is obviously the same as for the elliptic
$CMA$ obtained in \cite{bw}
\begin{equation}
\! X\! =\! i(\Omega_1\partial_2-\Omega_2\partial_1 -\Omega_{\bar
1}\partial_{\bar 2}+\Omega_{\bar 2}\partial_{\bar 1}) +C_1
(z^1\partial_1+\bar z^1\partial_{\bar 1}+u\partial_u) + iC_2
(z^2\partial_2-\bar z^2\partial_{\bar 2}) + H\partial_u \label{X}
\end{equation}
 where $C_1$ and $C_2$ are real constants and
$\Omega(z^i, \bar z^k)$ and $H(z^i, \bar z^k)$ are arbitrary
solutions of the linear system
\[ \Omega_{1\bar 1}=0, \quad \Omega_{2\bar 2}=0, \quad \Omega_{1\bar 2}=0,
\quad \Omega_{2\bar 1}=0 \] so that $\Omega =
\omega(z^i)+\bar\omega(\bar z^i)$ and $H = h(z^i)+\bar h(\bar
z^i)$. The corresponding symmetry characteristic \cite{olv} has
the form
\begin{eqnarray}
\widehat\eta & = & i(u_1\Omega_2-u_2\Omega_1+u_{\bar
2}\Omega_{\bar 1} -u_{\bar 1}\Omega_{\bar 2})
\label{sym}
\\
&& +C_1 (u-z^1u_1-\bar z^1u_{\bar 1}) -iC_2 (z^2u_2-\bar
z^2u_{\bar 2}) + H . \nonumber
\end{eqnarray}
Symmetries $\varphi$ and $\psi$ can be chosen as special cases of
the expression (\ref{sym}).

In this paper we shall restrict ourselves only to the case
$\psi=\varphi$ with the following two particular choices of
partner symmetries: combined translational symmetries in $z^1,
\bar z^1$ and $u$
\begin{equation}
\psi = \varphi = u_1 + u_{\bar 1} + h(z^2) + \bar h(\bar z^2)
\label{trans}
\end{equation}
and dilatational symmetries in the same variables
\begin{equation}
\psi = \varphi = u - z^1 u_1 - \bar z^1 u_{\bar 1}. \label{dilat}
\end{equation}

\section{Legendre transform of partner symmetries and ultra-hyperbolic metrics}
\setcounter{equation}{0} \label{sec-legendr-hcma}

Let us perform the Legendre transformation of equations
(\ref{hcma}) and (\ref{psi}) to new variables
\begin{equation}
v = u - z^1 u_1 - \bar z^1 u_{\bar 1},\quad p = u_1,\quad \bar p =
u_{\bar 1},\quad z^1 = -v_p,\quad \bar z^1 = -v_{\bar p}
\label{legendre}
\end{equation}
where $v=v(p,\bar p,z^2,\bar z^2)$. Then equation (\ref{hcma}) in
the new variables becomes
\begin{equation}
v_{p\bar p}v_{2\bar 2} - v_{p\bar 2}v_{\bar p 2} = v_{p\bar p}^2 -
v_{pp}v_{\bar p\bar p}
\label{leg_hcma}
\end{equation}
and equations (\ref{psi}) together with their complex conjugates
take the form
\begin{eqnarray}
& & \varphi_p v_{\bar p\bar p} - i\varphi_{\bar 2} v_{p\bar p} -
\varphi_{\bar p} (v_{p\bar p} - iv_{p\bar 2}) = 0 \nonumber
\\ & & \varphi_{\bar p} v_{pp} + i\varphi_{2} v_{p\bar p} -
\varphi_{p} (v_{p\bar p} + iv_{\bar p 2}) = 0 \label{v_eqn}
\\ & & \varphi_p \varphi_{\bar p} (2v_{p\bar p}- v_{2\bar 2}) -
(\varphi_{\bar p}^2 + i\varphi_{\bar p} \varphi_{\bar 2}) v_{pp} -
(\varphi_p^2 - i\varphi_p \varphi_2) v_{\bar p\bar p} \nonumber
\\ & & \mbox{} + (\varphi_2 \varphi_{\bar 2} + i\varphi_{\bar 2} \varphi_p
- i\varphi_2 \varphi_{\bar p}) v_{p\bar p} = 0  \nonumber
\end{eqnarray}
the last equation coinciding with its complex conjugate. Here we
keep the same notation $\varphi$ for the Legendre transform of the
symmetry characteristic which now depends on $p, \bar p$ instead
of $z^1, \bar z^1$.

Ultra-hyperbolic metric (\ref{metr}) governed by the field
equation (\ref{hcma}) after Legendre transformation
(\ref{legendre}) becomes
\begin{eqnarray}
& & ds^2 = \frac{1}{(v_{pp}v_{\bar p\bar p}-v_{p\bar
p}^2)}\biggl[v_{pp}(v_{p\bar p}dp+v_{\bar p2}dz^2)^2 + v_{\bar
p\bar p}(v_{p\bar p}d\bar p+v_{p\bar 2}d\bar z^2)^2 \nonumber
\\ & & \mbox{} + \frac{(v_{pp}v_{\bar p\bar p}+v_{p\bar p}^2)}{v_{p\bar
p}} \Bigl|v_{p\bar p}dp+v_{\bar p2}dz^2\Bigr|^2\biggr] -
\frac{(v_{pp}v_{\bar p\bar p}-v_{p\bar p}^2)}{v_{p\bar
p}}\,dz^2d\bar z^2
\label{legenmetr}
\end{eqnarray}
where we have used the Legendre transform (\ref{leg_hcma}) of
$HCMA$ in the last term.

The Legendre transforms of translational and dilatational
symmetries (\ref{trans}) and (\ref{dilat}) become respectively
\begin{equation}
\psi = \varphi = p + \bar p + h(z^2) + \bar h(\bar z^2)
\label{legtrans}
\end{equation}
and
\begin{equation}
\psi = \varphi = v.
\label{legdilat}
\end{equation}

 With the choice (\ref{legtrans}) equations
(\ref{v_eqn}) become linear
\begin{eqnarray}
& & v_{\bar p\bar p} - \left(i\bar h^\prime(\bar z^2) + 1\right)
v_{p\bar p} + iv_{p\bar 2} = 0 \nonumber
\\ & & v_{pp} + \left(ih^\prime(z^2) - 1\right) v_{p\bar p}
- iv_{\bar p 2} = 0
\label{vtrans}
\\ & & 2v_{p\bar p}- v_{2\bar 2} - \left(i\bar h^\prime(\bar z^2) + 1\right) v_{pp}
+ \left(ih^\prime(z^2) - 1\right) v_{\bar p\bar p}
 \nonumber
\\ & & \mbox{} + \left[h^\prime(z^2) \bar h^\prime(\bar z^2)
- i\left(h^\prime(z^2) - \bar h^\prime(\bar z^2)\right)\right]
v_{p\bar p} = 0. \nonumber
\end{eqnarray}
This system has no nontrivial differential compatibility
conditions since the equations $(v_{p\bar 2})_2=(v_{2\bar 2})_p$,
$(v_{\bar p 2})_{\bar 2}=(v_{2\bar 2})_{\bar p}$ and $(v_{p\bar
2})_{\bar p 2}=(v_{\bar p 2})_{\bar 2 p}$ are satisfied
identically.

With the choice (\ref{legdilat}) equations (\ref{v_eqn}) are still
non-linear
\begin{eqnarray}
& & v_p v_{\bar p\bar p} - i v_{\bar 2} v_{p\bar p} - v_{\bar
p}(v_{p\bar p} - iv_{p\bar 2}) = 0 \nonumber
\\ & & v_{\bar p} v_{pp} + i v_{2} v_{p\bar p} - v_{p}(v_{p\bar p}
+ iv_{\bar p 2}) = 0 \nonumber
\\ & & (v_2v_{\bar 2} + iv_{\bar 2}v_p - i v_2v_{\bar p} + 2v_pv_{\bar p}) v_{p\bar p}
 \nonumber
\\ & & \mbox{} - (v_{\bar p}^2 + iv_{\bar p} v_{\bar 2}) v_{pp} - (v_p^2 -
iv_pv_2) v_{\bar p\bar p} - v_pv_{\bar p} v_{2\bar 2} = 0
\label{vdilat}
\end{eqnarray}
again with no nontrivial differential compatibility conditions.
\begin{theorem}
Solutions of the over-determined system (\ref{vdilat}) are
functionally invariant, i.e. if $v$ is a solution to
(\ref{vdilat}), then $f(v)$ is also a solution to (\ref{vdilat})
whenever $f$ is an arbitrary function $\in C^2$.
\end{theorem}
The proof follows from a substitution of $f(v)$ instead of $v$ in
equations (\ref{vdilat}).

There is a particular class of solutions of the system
(\ref{vdilat}) satisfying a linear over-determined system of six
equations
\begin{eqnarray}
 & & v_{p\bar p} = a v_p + \bar a v_{\bar p} \nonumber
\\ & & v_{pp} = (\bar a + i\bar b) v_{p} - i\bar a v_{2}
\nonumber
\\ & & v_{\bar p\bar p} = (a - ib) v_{\bar p} + ia v_{\bar 2}
\label{lindilat}
\\ & & v_{p\bar 2} = b v_p - i \bar a v_{\bar p} + \bar av_{\bar 2}
\nonumber
\\ & & v_{\bar p 2} = \bar b v_{\bar p} + i a v_p + a v_2
\nonumber
\\ & & v_{2\bar 2} = (a + ib) v_p + bv_2 + (\bar a - i\bar b) v_{\bar p}
+ \bar b v_{\bar 2} \nonumber
\end{eqnarray}
where $a$ and $b$ are arbitrary complex constants. Here the first
three of these equations generate the last three equations as
their differential compatibility conditions.

The system (\ref{lindilat}) also has no nontrivial differential
compatibility conditions, i.e. equations $(v_{pp})_{\bar
p}=(v_{p\bar p})_p$\,, $(v_{p\bar 2})_2=(v_{2\bar 2})_p$ and their
complex conjugates are identically satisfied. The substitution of
the second derivatives from (\ref{lindilat}) to (\ref{vdilat})
identically satisfies the latter equations.

General solution of the system (\ref{lindilat}) is obtained in the
form
\begin{equation}
v = \sum_{j=1}^{n} C_j \exp{\Bigl(\alpha_j p + \bar\alpha_j\bar p
+ \beta_j z^2 + \bar\beta_j\bar z^2\Bigr)} \label{sol1}
\end{equation}
where the sum over integer $j$ may contain finite or infinite
number of terms, $C_j$ are arbitrary real constants and $\alpha_j$
and $\beta_j$ are determined by the equations
\begin{equation}
|\alpha_j|^2 = a \alpha_j + \bar a \bar\alpha_j ,
\label{alphadet}
\end{equation}
\begin{equation}
\beta_j = i \,\frac{\alpha_j^2-(\bar a+i \bar b)\alpha_j}{\bar a}
\label{betadet}
\end{equation}
together with the complex conjugate to (\ref{betadet}). The first
equation (\ref{alphadet}) is easily solved in polar coordinates
$a=re^{i\theta}$, $\bar a=re^{-i\theta}$, $\alpha=\chi e^{i\mu}$,
$\bar\alpha=\chi e^{-i\mu}$ in the form
\begin{equation}
\chi = 2r\cos{(\theta+\mu)}
\label{explicit}
\end{equation}
where $\mu$ remains as a free parameter.

Substituting for $v$ in (\ref{legenmetr}) any solution of the form
(\ref{sol1}) satisfying  (\ref{explicit}) and (\ref{betadet}) we
obtain explicitly an ultra-hyperbolic `heavenly' \cite{pleb}
metric. Since generic solution of the form (\ref{sol1}) obviously
depends on four independent variables and hence is non-invariant,
the resulting metric has no Killing vectors. The justification for
this statement is given in section \ref{sec-killing} where we
examine the Killing equations for (\ref{legenmetr}).

Though we have not found the general solution of the non-linear
system (\ref{vdilat}), we essentially enlarge the class of its
solutions using the property of functional invariance from
theorem\ \thetheorem. Due to this property any smooth function of
the solution (\ref{sol1}) is again a solution of (\ref{vdilat})
though not of the system (\ref{lindilat}) since the latter system
does not possess the above mentioned property. These more general
solutions also can be used in the formula (\ref{legenmetr}) giving
explicitly more classes of 4-dimensional ultra-hyperbolic heavenly
metrics admitting no Killing vectors.

To simplify the solution of the linear system (\ref{vtrans}), we
will restrict ourselves to the linear function $h(z^2)$: $h=\nu
z^2$, $\bar h = \bar\nu \bar z^2$ where $\nu$ is an arbitrary
constant, so that we obtain the linear system with constant
coefficients
\begin{eqnarray}
& & v_{\bar p\bar p} - \left(i\bar\nu + 1\right) v_{p\bar p} +
iv_{p\bar 2} = 0 \nonumber
\\ & & v_{pp} + \left(i\nu - 1\right) v_{p\bar p}
- iv_{\bar p 2} = 0
\label{constrans}
\\ & & \left[\, 2 + |\nu|^2 - i\left(\nu - \bar\nu\right)\right] v_{p\bar p} - \left(i\bar\nu + 1\right) v_{pp}
+ \left(i\nu - 1\right) v_{\bar p\bar p} - v_{2\bar 2}  = 0.
 \nonumber
\end{eqnarray}

We note that the particular solution (\ref{sol1}) of the system
(\ref{vdilat}) satisfying (\ref{lindilat}) turns out to be also a
particular solution of the system (\ref{constrans}) with the
additional constraint on the parameters in (\ref{lindilat})
\begin{equation}
\bar a = -a,\qquad b = (\bar\nu - i)a
\label{addit}
\end{equation}
where the second equation is just the relation between parameters
of the two systems. The constraints (\ref{explicit}) and
(\ref{betadet}) on the parameters in the solution (\ref{sol1})
then take the form ($\theta = \pi/2$)
\[\chi = -2r\sin{\mu},\quad \beta = -i\,\frac{\alpha^2}{a} + \nu . \]

It is easy to find general solution of the linear system
(\ref{constrans}) with constant coefficients in the form similar
to (\ref{sol1})
\begin{equation}
v = \sum_{j=1}^{n} C_j \exp{\Bigl(\alpha_j p + \bar\alpha_j\bar p
+ \beta_j z^2 + \bar\beta_j\bar z^2\Bigr)} \label{sol2}
\end{equation}
where the sum over integer $j$ may again contain finite or
infinite number of terms, $C_j$ are arbitrary real constants and
$\beta_j$ are expressed through $\alpha_j$ by the relation
\begin{equation}
\beta_j = \left(\nu + i -
i\,\frac{\alpha_j}{\bar\alpha_j}\right)\alpha_j \label{delgam}
\end{equation}
and its complex conjugate.

The substitution of any solution of the form (\ref{sol2}) with
(\ref{delgam}) for $v$ in the formula (\ref{legenmetr}) gives us
explicitly another class of ultra-hyperbolic heavenly metrics. The
solution (\ref{sol2}) depends on four variables in the generic
case, when the number of terms in the sum (\ref{sol2}) is not less
than four, and hence the corresponding metric has no Killing
vectors. We shall prove this in section \ref{sec-killing} by an
examination of the Killing equations.

\section{Symmetries and recursions of the second heavenly
equation of Pleba\~nski} \setcounter{equation}{0}
\label{sec-sym-heaven}

Let $\theta=\theta(w,z,x,y)$ be holomorphic complex-valued
function of four complex variables in some local coordinate system
on a complex $4$-dimensional manifold $\cal M$. The heavenly
metric of Pleba\~nski \cite{pleb} locally defined on $\cal M$ is
defined by (we skip here an overall factor $2$)
\begin{equation}
ds^2 = dw dx + dz dy - \theta_{xx}dz^2 - \theta_{yy}dw^2 +
2\theta_{xy}dw dz
\label{metr2}
\end{equation}
where subscripts denote partial derivatives with respect to
corresponding variables. This metric is governed by the second
heavenly equation \cite{pleb}
\begin{equation}
\theta_{xw}+\theta_{yz}+\theta_{xx}\theta_{yy}-\theta_{xy}^2 = 0
\label{heav2}
\end{equation}
for the potential $\theta$ in the metric (\ref{metr2}).

\subsection{Recursion relations from the divergence form of the determining
equation for symmetries} \label{sec-partner-heaven}

The determining equation for symmetries of the heavenly equation
(\ref{heav2}) has the form
\begin{equation}
\varphi_{xw}+\varphi_{yz}+\theta_{yy}\varphi_{xx}+\theta_{xx}\varphi_{yy}
-2\theta_{xy}\varphi_{xy} = 0
\label{defsym2}
\end{equation}
which is a linearization of (\ref{heav2}). It can also be written
as $\Box_\theta\varphi=0$ with the operator $\Box_\theta$ defined
by
\begin{equation}
\Box_\theta = D_xD_w + D_yD_z + \theta_{yy}D_x^2 +
\theta_{xx}D_y^2 - 2\theta_{xy}D_xD_y \label{delta2}
\end{equation}
where $D_x,D_w,D_y,D_z$ denote total derivatives with respect to
corresponding variables. We note that the determining equation
(\ref{defsym2}) can be presented in the form of total divergence,
i.e. differential conservation law
\begin{equation}
D_x(\varphi_w + \theta_{yy}\varphi_x - \theta_{xy}\varphi_y) +
D_y(\varphi_z + \theta_{xx}\varphi_y - \theta_{xy}\varphi_x) = 0
\label{div2}
\end{equation}
so that there locally exists a potential $\psi$ such that
\begin{equation}
\psi_y = \varphi_w + \theta_{yy}\varphi_x -
\theta_{xy}\varphi_y,\quad \psi_x = -(\varphi_z -
\theta_{xy}\varphi_x + \theta_{xx}\varphi_y)
\label{psi_fi}
\end{equation}
and differential compatibility conditions $(\psi_y)_x=(\psi_x)_y$
for the system (\ref{psi_fi}) coincide with the determining
equation for symmetries (\ref{defsym2}).

We introduce linear differential operators
\begin{equation}
L_y = D_w + \theta_{yy}D_x - \theta_{xy}D_y,\qquad L_x = - (D_z -
\theta_{xy}D_x + \theta_{xx}D_y) \label{Loper}
\end{equation}
so that the operator (\ref{delta2}) takes the form
\begin{equation}
\Box_\theta = D_xL_y - D_yL_x = L_yD_x - L_xD_y \label{de2L}
\end{equation}
and the relations (\ref{psi_fi}) become
\begin{equation}
\psi_y = L_y\varphi,\qquad \psi_x = L_x\varphi. \label{ps_L_fi}
\end{equation}
The commutator of the two operators has the form
\[ [L_x,L_y] =
(\theta_{xw}+\theta_{yz}+\theta_{xx}\theta_{yy}-\theta_{xy}^2)_yD_x
-(\theta_{xw}+\theta_{yz}+\theta_{xx}\theta_{yy}-\theta_{xy}^2)_xD_y\]
and hence $[L_x,L_y]=0$ on the solution manifold of the second
heavenly equation (\ref{heav2}). Alternatively, vanishing of the
commutator of $L_x$ and $L_y$ reproduces the equation
\[\theta_{xw}+\theta_{yz}+\theta_{xx}\theta_{yy}-\theta_{xy}^2 = C_w(w,z) \]
where $C(w,z)$ an arbitrary function of $w$ and $z$ which after
redefining $\theta$ to $\theta +xC(w,z)$ coincides with the
original equation (\ref{heav2}). Note that this redefinition does
not change the equations (\ref{defsym2}) - (\ref{ps_L_fi}).

Let $\varphi$ be a symmetry of (\ref{heav2}) so that it satisfies
the determining equation (\ref{defsym2}) $\Box_\theta\varphi=0$
and $\psi$ be a corresponding potential for $\varphi$ related to
it by formulas (\ref{ps_L_fi}). Then a simple calculation shows
that $\psi$ is also a symmetry, i.e. it satisfies the same
equation (\ref{defsym2})
\[\Box_\theta\psi = L_y\psi_x - L_x\psi_y = -[L_x,L_y]\varphi = 0 \]
where we have substituted $\psi_x$ and $\psi_y$ from
(\ref{ps_L_fi}) and used the commutativity of $L_x$ and $L_y$ on
the solution manifold of (\ref{heav2}). Therefore, the potential
$\psi$ for any symmetry $\varphi$ is also a symmetry and hence the
equations (\ref{ps_L_fi}) are {\it recursion relations} for
partner symmetries of (\ref{heav2}).

The Lax pair of Mason and Newman for the equation (\ref{heav2})
\cite{masnew,mw} can be expressed through $L_x$ and $L_y$ as
\begin{equation}
L_0 = D_y -\lambda L_y, \qquad L_1 = D_x -\lambda L_x \label{lax2}
\end{equation}
so that $[L_0,L_1]=\lambda^2[L_y,L_x]$ and the vanishing of the
commutator $[L_0,L_1]$ reproduces the equation (\ref{heav2}) up to
redefinition of $C(w,z)$, same as for $[L_x,L_y]=0$.

Recursion relations (\ref{ps_L_fi}) for symmetries can be
expressed in terms of non-local recursion operator $R$ defined by
\begin{equation}
\psi=R\varphi=D_y^{-1}L_y\varphi,\qquad
\psi=R\varphi=D_x^{-1}L_x\varphi. \label{R_oper}
\end{equation}

\subsection{Point symmetries of the second heavenly equation}
\label{sec-pointsym}

Here again we will avoid operating explicitly with non-local
symmetries and use instead the equations (\ref{psi_fi}) only with
the point partner symmetries $\varphi$ and $\psi$. Then these
equations are equivalent to an invariance condition for solutions
of the second heavenly equation with respect to non-local symmetry
which is a linear combination of $\psi$ and the non-local symmetry
generated from $\varphi$ by the recursion operator. Such solutions
will still be non-invariant in the usual sense.

Though here we use only an obvious translational symmetry for this
purpose, it is convenient for future analysis in the same
framework to present explicitly basis generators of the complete
symmetry algebra of point symmetries for the second heavenly
equation (\ref{heav2}). Symmetry generators of its one-parameter
subgroups have the form
\begin{eqnarray}
&\! & X_1 = \partial/\partial x , \quad X_2 = 2z\partial/\partial
x - xy\partial/\partial\theta ,\quad X_3 = y\partial/\partial y +
w\partial/\partial w + \theta\partial/\partial\theta \nonumber
\\ &\! & X_4 = x\partial/\partial x + y\partial/\partial y +
3\theta \partial/\partial\theta ,\quad Y_a = (ya_w-xa_z)
\partial/\partial\theta , \quad H_d = \partial/\partial\theta \nonumber
\\ &\! & Z_b = b_w \partial/\partial x + b_z \partial/\partial y +
(1/2)(x^2b_{zz} + y^2b_{ww} - 2xyb_{zw}) \partial/\partial\theta
\label{generat}
\\ &\! & G_c = (xc_{zw} - yc_{ww}) \partial/\partial x + (xc_{zz} -
yc_{zw}) \partial/\partial y + c_w \partial/\partial z - c_z
\partial/\partial w
\nonumber
\\ &\! &\mbox{} + (1/6)(x^3c_{zzz} - 3x^2yc_{zzw} + 3xy^2c_{zww} -
y^3c_{www})\partial/\partial\theta
\nonumber
\end{eqnarray}
where $a(z,w)$, $b(z,w)$, $c(z,w)$ and $d(z,w)$ are arbitrary
functions of $z, w$. Since some of the generators contain
arbitrary functions, the total symmetry group is an infinite Lie
(pseudo)group.

Of course, some of these symmetries are quite obvious, such as
translations in all independent variables and combined dilatations
\[ y' = \lambda y,\quad w' = \lambda w,\quad \theta' = \lambda
\theta\qquad {\rm and} \qquad x' = \lambda x,\quad y' = \lambda
y,\quad \theta' = \lambda^3 \theta \] generated by $X_3$ and $X_4$
respectively. From symmetry considerations one may wonder where is
the generator of the uniform dilatations in $x, z, \theta$:
$\tilde X_3 = x\partial/\partial x + z\partial/\partial z +
\theta\partial/\partial\theta$ ? The answer is that for a
particular choice $c=zw$ the generator $G_{zw}$ reduces to another
generator of dilatations
\[ G_{zw} = x\partial/\partial x - y\partial/\partial y
+ z\partial/\partial z - w\partial/\partial w \] so that $\tilde
X_3 = G_{zw} + X_3$. Seemingly missing generators of translations
in $y , z, w$ can also be found in (\ref{generat}) for certain
special choices of $b$ and $c$ as
\[ Z_z = \partial/\partial y , \quad G_w = \partial/\partial z ,
\quad G_{-z} = \partial/\partial w . \] We also have
the generator of simultaneous rotations in $(x,y)$ and $(w,z)$
complex planes
\[G_c = x\partial/\partial y - y\partial/\partial x
+ w\partial/\partial z - z\partial/\partial w \] which is obtained
from $G_c$ by the choice of $c = (z^2+w^2)/2$.

In the next section we shall use only one of the simplest
translational symmetries from (\ref{generat}), namely $G_z =
-\partial/\partial w$ with the characteristic $\theta_w$, and
still obtain very non-trivial results.

For completeness we present a table of commutators of the
generators (\ref{generat}) where the commutator $[X_i,X_j]$ stands
at the intersection of $i$th row and $j$th column. It is
convenient to introduce for this table the following shorthand
notation for a skew-symmetric differentiation of a pair of
functions $a(z,w)$ and $b(z,w)$: $a\wedge b = a_zb_w - b_za_w$ and
$s = xw + yz$, so that $xa_z - ya_w = a\wedge s$ and $x(f\wedge
b)_z - y (f\wedge b)_w = (f\wedge b)\wedge s$. We also denote
$\hat c_z = zc_z-c$ and $\hat c_w = wc_w-c$.
\begin{table}[ht]
\hspace{-1cm}
\begin{tabular}{|c|c|c|c|c|c|c|c|c|}
\hline    &$X_1$&$X_2$  &$X_3$& $X_4$ &$Y_a$       &$Z_b$        &$G_c$   &$H_d$
\\ \hline
    $X_1$ & $0$ & $-Y_w$& $0$ & $X_1$ &$-H_{a_z}$  &$-Y_{b_z}$   &$Z_{c_z}$ &$0$
\\ \hline
    $X_2$ &$Y_w$& $0$ & $0$   & $X_2$ &$-2H_{za_z}$&$-Y_{2\hat b_z+b}$&$2Z_{\hat c_z}$&$0$
\\ \hline
    $X_3$ &$0$  &$0$  & $ 0$  &$0$  &$-H_{(wa_w)\wedge s}$&$Z_{\hat b_w}$&$G_{\hat c_w}$&$H_{\hat d_w}$
\\ \hline
    $X_4$ &$-X_1$&$-X_2$&$0$  & $0$ &$2H_{a\wedge s}$&$-Z_b$&$0$      &$-3H_d$
\\ \hline
    $Y_e$ &$H_{e_z}$&$2H_{ze_z}$&$H_{(we_w)\wedge s}$&$-2H_{e\wedge s}$&$0$&$H_{e\wedge b}$&$H_{(e\wedge c)\wedge s}$&$0$
\\ \hline
    $Z_f$ &$Y_{f_z}$&$Y_{2\hat f_z+f}$&$-Z_{\hat f_w}$&$Z_f$&$-H_{a\wedge f}$&$-H_{(f\wedge b)\wedge s}$&$Z_{c\wedge f}$&$0$
\\ \hline
    $G_g$ &$-Z_{g_z}$&$-2Z_{\hat g_z}$&$-G_{\hat g_w}$&$0$&$-H_{(a\wedge g)\wedge s}$&$-Z_{g\wedge b}$&$G_{c\wedge g}$&$H_{d\wedge g}$
\\ \hline
    $H_h$ &$0$   &$0$    &$-H_{\hat h_w}$&$3H_h$ &$0$ &$0$  &$-H_{h\wedge c}$& $0$
\\ \hline
\end{tabular}
\caption{Commutators of point symmetries of the second heavenly
equation.}
\end{table}
In table\,\thetable\ $a,b,c,d,e,f,g,h$ are arbitrary functions of
$z,w$.

 All contact symmetries of the second heavenly equation
(\ref{heav2}) coincide with its prolonged point symmetries.

Symmetries of the second heavenly equation were also studied in
\cite{lp,dm2000}.

\subsection{Legendre transformation and heavenly\\ metric}
\label{sec-legendr-heaven}

Our final goal is to end up with linear equations. This will be
achieved in the next section by applying the partial Legendre
transformation
\begin{equation}
u=\theta-w\theta_w-y\theta_y,\quad \theta_w=t,\quad
\theta_y=r,\quad w=-u_t,\quad y=-u_r \label{legendr}
\end{equation}
to second heavenly equation (\ref{heav2}) together with equations
(\ref{psi_fi}) for some choices of local symmetries $\varphi$ and
$\psi$ . The existence condition for Legendre transformation
(\ref{legendr}) has the form
\begin{equation}
\Delta = u_{tt}u_{rr} - u_{rt}^2  \ne 0.
\label{delta}
\end{equation}
The Legendre transform of (\ref{heav2}) is
\begin{equation}
u_{tt}(u_{xx}+u_{rz})+u_{xt}(u_{rr}-u_{xt})-u_{rt}(u_{rx}+u_{tz}) = 0.
\label{legheav2}
\end{equation}
The Legendre
transformation (\ref{legendr}) of the metric (\ref{metr2}) results in
\begin{eqnarray}
& & ds^2 = \frac{[u_{tt}(u_{tt}dt+u_{tr}dr+u_{tx}dx+u_{tz}dz) +
(u_{tt}u_{rx} - u_{tr}u_{tx})dz]^2}{u_{tt}(u_{tt}u_{rr} -
u_{tr}^2)} \nonumber
\\ & & \mbox{} - \frac{(u_{tt}u_{xx} - u_{tx}^2)}{u_{tt}}\,dz^2 - (u_{tt}dt + u_{tr}dr + u_{tx}dx +
u_{tz}dz)dx \nonumber
\\ & & \mbox{} - (u_{rt}dt + u_{rr}dr + u_{rx}dx + u_{rz}dz)dz
\label{legmetr2}
\end{eqnarray}
with the potential $u$ satisfying the equation (\ref{legheav2}).

\section{Example: use of translational symmetries}
\setcounter{equation}{0} \label{sec-translsym}
\subsection{Case of equal symmetries}
\label{sec-equalsym}

At first we consider the case when the two partner symmetries are
equal to each other $\varphi =\psi$ and we choose $\varphi$ to be
equal to the characteristic of the translational symmetry
$\varphi=\theta_w$. Then the equations (\ref{psi_fi}) take the
form
\begin{eqnarray}
-\theta_{yw}+\theta_{ww}+\theta_{yy}\theta_{wx}-\theta_{xy}\theta_{wy}=0
\label{tr_tr1}
\\ \theta_{xw}+\theta_{wz}+\theta_{xx}\theta_{wy}-\theta_{xy}\theta_{wx}=0
\label{tr_tr2}
\end{eqnarray}
so that together with (\ref{heav2}) we obtain a system of three
equations.

After applying Legendre transformation (\ref{legendr}) equation
(\ref{tr_tr1}) is linearized in the form
\begin{equation}
u_{rt}+u_{rr}-u_{xt} = 0 \label{lin1}
\end{equation}
and with the aid of the equation (\ref{lin1}) the equations
(\ref{tr_tr2}), (\ref{heav2}) become respectively
\begin{equation}
u_{rt}(u_{xx}+u_{rz})-u_{rr}(u_{rx}+u_{xt}+u_{tz})=0,
\label{tr2}
\end{equation}
\begin{equation}
-u_{tt}(u_{xx}+u_{rz})+u_{rt}(u_{rx}+u_{xt}+u_{tz})=0.
\label{lh2}
\end{equation}
Solving the system (\ref{tr2}), (\ref{lh2}) algebraically we
obtain two linear equations
\begin{equation}
u_{xx}+u_{rz} = 0,
\label{lin2}
\end{equation}
\begin{equation}
u_{rx}+u_{xt}+u_{tz} = 0
\label{lin3}
\end{equation}
since the determinant of this system is non-zero due to the
condition $\Delta\ne 0$ with $\Delta$ defined in (\ref{delta}).

The system of the three linear equations (\ref{lin1}),
(\ref{lin2}) and (\ref{lin3}) corresponds to the original second
heavenly equation (\ref{heav2}) plus two differential constraints
(\ref{tr_tr1}) and (\ref{tr_tr2}).

Solution of this linear system has the form
\begin{equation}
u = \sum_{j=1}^n C_j \exp{\Bigl(\alpha_j t + \beta_j r + \gamma_j x +
\delta_j z\Bigr)}
\label{sol3}
\end{equation}
where $C_j$ are arbitrary constants and the parameters satisfy the
relations
\begin{equation}
\alpha_j = \frac{\beta_j^2}{\gamma_j - \beta_j}\,,\quad \delta_j =
- \frac{\gamma_j^2}{\beta_j}\,.
\label{alp_del}
\end{equation}

Substitution of this solution to the Legendre transform
(\ref{legmetr2}) of the heavenly metric (\ref{metr2}) gives an
explicit form of such a metric. The solution (\ref{sol3}) depends
on four variables in the generic case, when the number of terms in
the sum (\ref{sol3}) is greater than three, and hence the
corresponding metric has no Killing vectors.

\subsection{Solutions invariant with respect to higher\\ symmetry}
\label{sec-highsym}

Here we consider the case when $\varphi$ is the translational
symmetry $\varphi=\theta_w$ and its partner symmetry
characteristic $\psi$ is equal to zero $\psi=0$. This means the
invariance of the solution for $\theta$ with respect to the
nonlocal higher symmetry $\psi$ generated from $\varphi=\theta_w$
by the recursion relations (\ref{psi_fi})
\begin{equation}
L_y\varphi =
\theta_{ww}+\theta_{yy}\theta_{wx}-\theta_{xy}\theta_{wy}=0
\label{inv1}
\end{equation}
\begin{equation}
-L_x\varphi =
\theta_{wz}-\theta_{xy}\theta_{wx}+\theta_{xx}\theta_{wy} = 0
\label{inv2}
\end{equation}
so that we have again the system of three equations (\ref{inv1}),
(\ref{inv2}) and (\ref{heav2}).

Next we perform the partial Legendre transformation
(\ref{legendr}) of this system. The Legendre transform of
(\ref{inv1}) is
\begin{equation}
u_{rr} - u_{tx} = 0 . \label{lin4}
\end{equation}
With the use of (\ref{lin4}) the Legendre transforms of
(\ref{inv2}) and (\ref{heav2}) become respectively
\begin{equation}
u_{rr}(u_{rx}+u_{tz}) - u_{tr}(u_{rz}+u_{xx}) = 0 \label{leg2}
\end{equation}
\begin{equation}
u_{rt}(u_{rx}+u_{tz}) - u_{tt}(u_{rz}+u_{xx}) = 0 . \label{leg3}
\end{equation}
Due to the condition $\Delta\ne 0$ in (\ref{delta}) equations
(\ref{leg2}) and (\ref{leg3}) become
\begin{equation}
u_{rx}+u_{tz} = 0 \label{lin5}
\end{equation}
\begin{equation}
u_{rz}+u_{xx} = 0 \label{lin6}
\end{equation}
so that we end up with the system of linear equations
(\ref{lin4}), (\ref{lin5}) and (\ref{lin6}).

Solution of this linear system has again the form (\ref{sol3}) but
with the modified relations between parameters
\begin{equation}
\alpha_j = \frac{\beta_j^2}{\gamma_j}\,,\quad \delta_j = -
\frac{\gamma_j^2}{\beta_j}\,. \label{alp_del2}
\end{equation}

The corresponding explicit form of heavenly metric is obtained by
a substitution of the solution (\ref{sol3}) for $u$ into the
formula (\ref{legmetr2}). This metric generically has no Killing
vectors for the same reason as mentioned above.

\section{Relationship between Killing vectors and symmetries of
the potential} \label{sec-killing} \setcounter{equation}{0}

Here we shall give proof of the non-existence of any Killing
vectors for our metrics.

Our emphasis will be on the {\it existence} problem. Thus we shall
make use of the fact that a vector is an invariant object and the
existence of a vector field in one frame implies its existence in
any other frame. We have started out with the simple looking
K\"ahler metric with the complicated $CMA$ equation as the
condition for Ricci-flatness and after a Legendre transformation
arrived at a complicated form of the metric with linear field
equations which are easily solved. With the Killing vector there
is a similar situation. Killing's equations are simple in the
K\"ahler form of the metric and we obtain a linear first-order
PDE, see (\ref{1stord}) below, which encodes all the information
in the Killing equations. On solutions of the field equations the
Legendre transformation induces point transformations between the
coordinates entering into the metric. This is a linear homogeneous
transformation between the components of the Killing vector and
its non-existence in one frame will imply its non-existence in any
other frame.

\subsection{Analysis of the Killing equations for the K\"ahler metric}

Thus we consider first the K\"ahler metric (\ref{metr}) and let
\begin{equation}
\vec v = \xi^k(z,\bar z)\frac{\partial}{\partial z^k} + \xi^{\bar
k}(z,\bar z)\frac{\partial}{\partial\bar z^{k}} \label{kill}
\end{equation}
be the Killing vector for (\ref{metr}) where $z=(z^1,z^2)$ and
$\bar z=(\bar z^1,\bar z^2)$ and summation over dummy indices
ranges over two values for both barred and unbarred indices. The
reality condition for $\vec v$ implies $\xi^{\bar k} =
\bar{\xi^k}$.

The Killing equations for the metric (\ref{metr}) fall into two
sets
\begin{equation}
u_{i\bar k} \, \xi^{\bar k}_{\,j} = 0 , \quad u_{k\bar
j}\,\xi^{k}_{\; \bar i} = 0 \label{1st}
\end{equation}
and
\begin{equation}
(\xi^k u_{k\bar j})_i + (\xi^{\bar k} u_{\bar k i})_{\bar j} = 0
\label{2nd}
\end{equation}
where subscripts denote partial derivatives. The determinant of
the linear equations (\ref{1st}) is non-zero due to the $CMA$
equation (\ref{ecma})
\[ {\rm det}(u_{i\bar k}) = u_{1\bar 1}u_{2\bar 2}
- u_{2\bar 1}u_{1\bar 2} = \pm 1\] and hence these equations have
only vanishing solutions $\xi^{\bar k}_j=0$, $\xi^k_{\bar i}=0$,
so that
\begin{equation}
\xi^i = \xi^i(z),\quad \xi^{\bar i} = \xi^{\bar i}(\bar z).
\label{holom}
\end{equation}
The remaining Killing equations become
\[ \left(\xi^k u_k + \xi^{\bar k} u_{\bar k}\right)_{i\,\bar j} = 0 \]
with the solution
\begin{equation}
u_k \xi^k(z) + u_{\bar k} \xi^{\bar k}(\bar z) = h(z) + \bar
h(\bar z) \label{1stord}
\end{equation}
where $h$ is an arbitrary biholomorphic function.

Hence the Killing equations are equivalent to the linear equation
(\ref{1stord}) for $\xi^k(z), \xi^{\bar k}(\bar
z)$ which should be satisfied for a given solution $u(z,\bar z)$ of $CMA$
if a Killing vector for the corresponding metric (\ref{metr}) exists.

We note that if $\xi^k, \xi^{\bar k}$ are chosen as coefficients
of a generator of point symmetries (\ref{X}) with $C_1 = 0$, then
(\ref{1stord}) coincides with the invariance condition
$\widehat\eta = 0$ for solutions of $CMA$, where $\widehat\eta$ is
the symmetry characteristic (\ref{sym}). More generally (\ref{1stord})
determines conditionally invariant solutions of $CMA$ \cite{LW}, if conditional
symmetries exist. Thus, any such
symmetry in a solution of $CMA$ implies the existence of a Killing
vector for the K\"ahler metric (\ref{metr}).

The Legendre transformation (\ref{legendre}) induces an invertible
point coordinate transformation
\begin{equation}
p = u_1(z,\bar z),\quad \bar p = u_{\bar 1}(z,\bar z),\quad
z^1 = -v_p(p,\bar p,z^2,\bar{z}^2),\quad
\bar{z}^1 = -v_{\bar p}(p,\bar p,z^2,\bar{z}^2)
\label{point}
\end{equation}
on solutions $u(z,\bar z)$ and $v(p,\bar p,z^2,\bar{z}^2)$ of $HCMA$
(\ref{hcma}) and its  Legendre transform (\ref{leg_hcma}) respectively
and $z^2, \bar{z}^2$ are not transformed. Under this transformation
the condition (\ref{1stord}), equivalent to the Killing equations,
results in
\begin{eqnarray}
& & p\,\xi^1(-v_p\,,z^2) + v_2\,\xi^2(-v_p\,,z^2) + \bar
p\,\xi^{\bar 1}(-v_{\bar p}\,,\bar{z^2}) + v_{\bar 2}\,\xi^{\bar
2}(-v_{\bar p}\,,\bar{z^2})\nonumber
\\ & & = h(-v_p\,,z^2) + \bar h(-v_{\bar p}\,,\bar{z^2})
\label{leg1st}
\end{eqnarray}

We should account also for a transformation of components of the Killing
vectors induced by the transformation (\ref{point}). Let
\begin{equation}
\vec v = \eta^p \frac{\partial}{\partial p} + \eta^{\bar p}
\frac{\partial}{\partial \bar p} + \eta^2 \frac{\partial}{\partial
z^2} + \eta^{\bar 2} \frac{\partial}{\partial \bar{z}^2}
\label{legkill}
\end{equation}
be the Killing vector (\ref{kill}) in the frame transformed by (\ref{point}).
The transformation law for the components of the Killing vector
is given by
\begin{eqnarray}
& & \xi^1 = -\left(\eta^p v_{pp} + \eta^{\bar p} v_{p\bar p} +
\eta^2 v_{p2} + \eta^{\bar 2} v_{p\bar 2} \right), \quad \xi^2 =
\eta^2 \nonumber
\\ & & \xi^{\bar 1} = -\left(\eta^p v_{\bar pp} + \eta^{\bar p} v_{\bar p\bar p} + \eta^2
v_{\bar p2} + \eta^{\bar 2} v_{\bar p\bar 2} \right), \quad
\xi^{\bar 2} = \eta^{\bar 2}
\label{transkill}
\end{eqnarray}
where the arguments of $\eta^p, \eta^2, \eta^{\bar p}, \eta^{\bar
2}$ and second derivatives of $v$ on the right hand sides of
(\ref{transkill}) consist of $ p, \bar p, z^2,
\bar z^2$. Hence the existence of $\vec v$ defined by (\ref{kill}) is
equivalent to that of $\vec v$ defined by (\ref{legkill}).
Given some solution $v$ of (\ref{leg_hcma}), a Killing vector for
the transformed metric (\ref{legenmetr}) exists only if one can satisfy
(\ref{leg1st}).

We need to check if this equation can be satisfied by our
solutions (\ref{sol1}) and (\ref{sol2}). Both of them have the
form
\begin{equation}
v = \sum_{j=1}^n C_j e^{\Phi_j}
\label{solform}
\end{equation}
where $C_j$ are arbitrary real constants,
\begin{equation}
\Phi_j = \alpha_j p +\bar\alpha_j \bar p +\beta_j z^2 + \bar\beta_j \bar{z^2}
\label{phase}
\end{equation}
and the parameters $\alpha_j,\beta_j$ satisfy the conditions (\ref{alphadet}),
(\ref{betadet}) for (\ref{sol1}) and (\ref{delgam}) for
(\ref{sol2}).

Let $n \ge 4$ in (\ref{solform}) and $\Phi_1,\Phi_2,\Phi_3,\Phi_4$
be linearly independent, {\it i.e.} the transformation
(\ref{phase}) from $p,\bar p,z^2,\bar{z^2}$ to
$\Phi_1,\Phi_2,\Phi_3,\Phi_4$ is invertible, provided that
$\alpha_1\alpha_2\alpha_3\alpha_4 \neq 0$ and the determinant of
the matrix of coefficients of (\ref{phase}) is non-zero
\begin{equation}
\left|
\begin{array}{cccc}
1 & e^{-2i\mu_1} & e^{2i\mu_1} & e^{-4i\mu_1}
\\ 1 & e^{-2i\mu_2} & e^{2i\mu_2} & e^{-4i\mu_2}
\\ 1 & e^{-2i\mu_3} & e^{2i\mu_3} & e^{-4i\mu_3}
\\ 1 & e^{-2i\mu_4} & e^{2i\mu_4} & e^{-4i\mu_4}
\end{array}\right|
\neq 0 ,
\label{cond1}
\end{equation}
where $\mu_j$ are the phases of $\alpha_j$. The same condition
holds for both solutions (\ref{sol1}) and (\ref{sol2}). Then
$p,\bar p,z^2,\bar{z}^2$ can be expressed through
$\Phi_1,\Phi_2,\Phi_3,\Phi_4$ and the same for
$\Phi_5,\ldots,\Phi_n$, so that $\Phi_j$ for $j=1,2,3,4$ can be
chosen as new independent variables in (\ref{leg1st}) and the
equation (\ref{leg1st}) takes the form
\begin{equation}
G(\Phi_1,\Phi_2,\Phi_3,\Phi_4) = 0.
\label{form2nd}
\end{equation}

Obviously the equation (\ref{leg1st}) cannot be satisfied
identically for any solution $v$ of (\ref{leg_hcma}) just by a
suitable choice of of functions $\xi^i,\xi^{\bar i},h,\bar h$
because of $p,\bar p,v_2,v_{\bar 2}$ entering explicitly its
coefficients. A similar remark applies also to the original
equation (\ref{1stord}) where generically the coefficients
$u_i,u_{\bar i}$ depend on $z$ and $\bar z$ together while the
unknowns $\xi^i,\xi^{\bar i}$ and $h,\bar  h$ depend only on $z$
or $\bar z$ separately. Hence (\ref{1stord}) and (\ref{leg1st})
should be considered as the equations determining particular
solutions of $CMA$ or (\ref{leg_hcma}) respectively for any choice
of  $\xi^i,\xi^{\bar i},h,\bar  h$ and this choice is constrained
by compatibility conditions of (\ref{1stord}) with $CMA$ and
(\ref{leg1st}) with (\ref{leg_hcma}).

In our case solutions of (\ref{leg_hcma}) are already determined
in (\ref{solform}) up to arbitrary constants by solving
second-order linear equations together with the
Legendre-transformed $HCMA$ and hence, having no functional
arbitrariness, they cannot satisfy in addition the first-order
equation (\ref{leg1st}). Thus for any choice of $\xi^i,\xi^{\bar
i},h,\bar h$ (\ref{leg1st}) is not an identity but an equation of
the form (\ref{form2nd}). This implies a dependence of the
independent variables which is a contradiction that proves
nonexistence of the Killing vectors for the metric
(\ref{legenmetr}) where the potential $v$ is determined by
(\ref{sol1}) and (\ref{sol2}) with $n\ge 4$ and the condition
(\ref{cond1}) is satisfied.

If $n\le 3$ then this reasoning obviously does not work and Killing vectors
may exist.

Thus we have proved the following theorem.
\begin{theorem}
Metric (\ref{legenmetr}) with $v$ defined either by (\ref{sol1})
with the conditions (\ref{alphadet}), (\ref{betadet}), or by
(\ref{sol2}) with the conditions (\ref{delgam}) provided that $n\ge
4$, $\alpha_1\alpha_2\alpha_3\alpha_4 \neq 0$ and satisfying the
non-degeneracy condition (\ref{cond1}) admits no Killing vectors.
\end{theorem}

\subsection{Analysis of the Killing equations for the second heavenly metric}

Now we shall perform a similar analysis for the heavenly metric
(\ref{metr2}) governed by the potential satisfying the second
heavenly equation (\ref{heav2}). Let
\begin{equation}
\vec\Omega = \xi^x(x,y,z,w)\frac{\partial}{\partial x} +
\xi^y(x,y,z,w)\frac{\partial}{\partial y} +
\xi^z(x,y,z,w)\frac{\partial}{\partial z} +
 \xi^w(x,y,z,w)\frac{\partial}{\partial w}
\label{kill2}
\end{equation}
denote the Killing vector for the metric (\ref{metr2}). The Killing equations
for this metric fall into three sets
\begin{eqnarray}
& & \xi^w_x = 0,\quad \xi^z_y = 0,\quad \xi^z_x+\xi^w_y = 0,\quad
\xi^y_x+\xi^w_z-2\theta_{xx}\xi^z_x = 0
\label{killeq1}
\\ & & \xi^x_x+\xi^w_w+2\theta_{xy}\xi^z_x = 0,\quad
\xi^y_y+\xi^z_z+2\theta_{xy}\xi^w_y = 0, \quad
\xi^x_y+\xi^z_w-2\theta_{yy}\xi^w_y = 0
\nonumber
\\[2mm] & & \xi^x\theta_{xxx}+\xi^y\theta_{yxx}+\xi^z\theta_{zxx}
+\xi^w\theta_{wxx} = \xi^y_z-2\theta_{xx}\xi^z_z+2\theta_{xy}\xi^w_z
\nonumber
\\ & & \xi^x\theta_{xyy}+\xi^y\theta_{yyy}+\xi^z\theta_{zyy}
+\xi^w\theta_{wyy} = \xi^x_w+2\theta_{xy}\xi^z_w-2\theta_{yy}\xi^w_w
\label{killeq2}
\\[2mm] & & 2\left(\xi^x\theta_{xxy}+\xi^y\theta_{yxy}+\xi^z\theta_{zxy}
+\xi^w\theta_{wxy}\right) + 2\theta_{xy}\left(\xi^z_z+\xi^w_w\right)
\nonumber
\\ & & -2\theta_{xx}\xi^z_w-2\theta_{yy}\xi^w_z+\xi^y_w+\xi^x_z = 0
\label{killeq3}
\end{eqnarray}
of seven, two and one equations respectively. The first subsystem
(\ref{killeq1}) is easily integrated to give
\begin{eqnarray}
& & \xi^x = -2a\theta_y-xd_w-yb_w-e,\quad \xi^y = 2a\theta_x-xd_z-yb_z+c
\nonumber
\\ & & \xi^z = ax+b,\quad \xi^w = -ay+d
\label{int1}
\end{eqnarray}
where $a$ is an arbitrary constant and $b,c,d,e$ are arbitrary functions
of $z,w$ only. The two equations (\ref{killeq2}) are integrated with
respect to $x$ and $y$ respectively and the results are substituted into
(\ref{killeq3}) which determines $y$- and $x$-dependences of the integration
"constants". Then we solve two remaining equations (\ref{killeq2})
algebraically with respect to $(\xi^x\theta_x+\xi^y\theta_y+\xi^z\theta_z
+\xi^w\theta_w)_x$ and $(\xi^x\theta_x+\xi^y\theta_y+\xi^z\theta_z
+\xi^w\theta_w)_y$ and then equate their cross-derivatives in $y$ and $x$
which gives additional constraints on the right-hand sides of these
equations. In particular, we obtain that
\begin{equation}
b = q_w+kz,\quad d = -q_z+kw
\label{bd}
\end{equation}
where $k$ is an arbitrary constant and $q$ is an arbitrary function of $z,w$.

For the purpose of further integration we present the second heavenly equation
(\ref{heav2}) in a divergence form
\begin{equation}
(\theta_y\theta_{xx}-\theta_x\theta_{yx}+2\theta_z)_y =
(\theta_y\theta_{xy}-\theta_x\theta_{yy}-2\theta_w)_x
\label{diver}
\end{equation}
so that we can locally define the potential $V$
\begin{equation}
V_x = \theta_y\theta_{xx}-\theta_x\theta_{yx}+2\theta_z,\quad
V_y = \theta_y\theta_{xy}-\theta_x\theta_{yy}-2\theta_w
\label{pot}
\end{equation}
up to an arbitrary term depending only on $z, w$. Then we can
further integrate the two equations
(\ref{killeq2}), whose left-hand sides after the first integration were
$(\xi^x\theta_x+\xi^y\theta_y+\xi^z\theta_z
+\xi^w\theta_w)_x$ and $(\xi^x\theta_x+\xi^y\theta_y+\xi^z\theta_z
+\xi^w\theta_w)_y$,  with respect to $x$ and $y$ which results in
\begin{eqnarray}
& & \xi^x\theta_x+\xi^y\theta_y+\xi^z\theta_z +\xi^w\theta_w
\equiv [x(q_{zw}-k)-yq_{ww}-e]\theta_x \nonumber
\\ & & + [xq_{zz}-y(q_{zw}+k)+c]\theta_y +(ax+kz+q_w)\theta_z
+ (kw-ay-q_z)\theta_w \nonumber
\\ & & = 2(aV-k\theta) +\frac{1}{6}\,(x^3q_{zzz}-3x^2yq_{zzw}
+3xy^2q_{zww}-y^3q_{www}) \nonumber
\\ & & + \frac{1}{2}\,[x^2c_z+xy(e_z-c_w)-y^2e_w] + x\rho
+y\sigma +\kappa
\label{1stint}
\end{eqnarray}
where we have used the expressions (\ref{int1}) for $\xi^x, \xi^y,
\xi^z, \xi^w$ in the left-hand side of this equation and $\rho,
\sigma, \kappa$ are new arbitrary functions of $z, w$.
Hence the Killing equations (\ref{killeq1}) - (\ref{killeq3}) are
equivalent to the linear first-order PDE (\ref{1stint}) which should
be identically satisfied for a given solution of the heavenly equation
(\ref{heav2}) with a suitable choice of arbitrary functions $q,c,e,\rho,
\sigma,\kappa$ of the variables $z,w$, arbitrary constants $a,k$ and the
integration "constant" for the potential $V$ in (\ref{pot}) depending on
$z,w$. If (\ref{1stint}) is satisfied then the components of the Killing
vector are given by (\ref{int1}) together with (\ref{bd}).

The Legendre transformation (\ref{legendr}) induces an invertible
point coordinate transformation
\begin{equation}
r = \theta_y(x,y,z,w),\; t = \theta_w(x,y,z,w),\quad
y = -u_r(x,r,z,t),\; w = -u_t(x,r,z,t)
\label{point2}
\end{equation}
on solutions $\theta(x,y,z,w)$ and $u(x,r,z,t)$ of the second heavenly
equation (\ref{heav2}) and its  Legendre transform (\ref{legheav2})
respectively and $x, z$ are not transformed. Let
\begin{equation}
\vec\Omega = \eta^x\frac{\partial}{\partial x}
+ \eta^r\frac{\partial}{\partial r} + \eta^z\frac{\partial}{\partial z}
+ \eta^t\frac{\partial}{\partial t}
\label{Om}
\end{equation}
be a Killing vector (\ref{kill2}) in the frame transformed by (\ref{point2}).
The components of the Killing vector are accordingly transformed by
\begin{eqnarray}
& & \xi^y = -\left(\eta^xu_{rx}+\eta^ru_{rr}+\eta^zu_{rz}+\eta^tu_{rt}\right),
\quad \xi^x = \eta^x \nonumber
\\ & & \xi^w = -\left(\eta^xu_{tx}+\eta^ru_{tr}+\eta^zu_{tz}+\eta^tu_{tt}\right)
, \quad \xi^z = \eta^z
\label{transkill2}
\end{eqnarray}
where the arguments of $\eta^x, \eta^r, \eta^z, \eta^t$ and second derivatives
of $u$ on the right-hand sides of (\ref{transkill2}) consist of $x,r,z,t$.

The Legendre transformation of the equation (\ref{1stint}) results in
\begin{eqnarray}
& & [x(q_{zw}-k)+u_rq_{ww}-e]u_x + r[xq_{zz}+u_r(q_{zw}-k)+c]
\nonumber
\\ & & +(ax+kz+q_w)u_z + t(-3ku_t+au_r-q_z) = 2(aP-ku) \nonumber
\\ & & + (1/6)(x^3q_{zzz}+3x^2u_rq_{zzw}
+3xu_r^2q_{zww}+u_r^3q_{www}) \nonumber
\\ & & + (1/2)[x^2c_z-xu_r(e_z-c_w)-u_r^2e_w] + x\rho
-u_r\sigma +\kappa
\label{1stint2}
\end{eqnarray}
where $u$ is a solution of the Legendre-transformed second heavenly
equation in (\ref{legheav2}), $w=-u_t(x,r,z,t)$ should be subsituted in
$q,c,e$, their derivatives, $\rho,\sigma$ and $\kappa$ while the transformed
potential
\[ P(x,r,z,t)=V\Bigl(x,-u_r(x,r,z,t),z,-u_t(x,r,z,t)\Bigr)\]
is determined by the transformed equations (\ref{pot}) defining the potential
$V$
\begin{eqnarray}
& & u_{tt}P_x - u_{tx}P_t = 2t(u_{rx}u_{tt}-u_{rt}u_{tx})
+ r(u_{tt}u_{xx}-u_{tx}^2) + 2u_zu_{tt} \nonumber
\\ & & u_{tt}P_r - u_{rt}P_t = 2t(u_{rr}u_{tt}-u_{rt}^2)
+ r(u_{rx}u_{tt}-u_{rt}u_{tx}) - u_xu_{tt}
\label{potP}
\end{eqnarray}
with the only differential compatibility condition coinciding with
(\ref{legheav2}).

We need to check if the equation (\ref{1stint2}) can be  satisfied
for our solutions (\ref{sol3}). They have again the form
(\ref{solform}) where
\begin{equation}
\Phi_j = \alpha_jt+\beta_jr+\gamma_jx+\delta_jz
\label{fi2}
\end{equation}
$C_j$ are arbitrary complex constants and the parameters satisfy
(\ref{alp_del}) or (\ref{alp_del2}).

Let $n \ge 4$ in (\ref{solform}) and $\Phi_1,\Phi_2,\Phi_3,\Phi_4$
be linearly independent, {\it i.e.} the transformation (\ref{fi2})
from $t,r,x,z$ to $\Phi_1,\Phi_2,\Phi_3,\Phi_4$ is invertible,
provided that $\beta_1\beta_2\beta_3\beta_4 \neq 0$ and the
determinant of the matrix of coefficients of (\ref{fi2}) is
non-zero. For the solution (\ref{sol3}) with the parameters
satisfying (\ref{alp_del}) this condition is
\begin{equation}
\left|
\begin{array}{cccc}
   \beta_1/(\gamma_1-\beta_1) & 1 &\gamma_1/\beta_1 &-\gamma_1^2/\beta_1^2
\\ \beta_2/(\gamma_2-\beta_2) & 1 &\gamma_2/\beta_2 &-\gamma_2^2/\beta_2^2
\\ \beta_3/(\gamma_3-\beta_3) & 1 &\gamma_3/\beta_3 &-\gamma_3^2/\beta_3^2
\\ \beta_4/(\gamma_4-\beta_4) & 1 &\gamma_4/\beta_4 &-\gamma_4^2/\beta_4^2
\end{array}\right|
\neq 0 .
\label{cond2}
\end{equation}
For the solution (\ref{sol3}) with the parameters satisfying (\ref{alp_del2})
the corresponding invertibility condition has the form
\begin{equation}
\left|
\begin{array}{cccc}
   \beta_1/\gamma_1 & 1 &\gamma_1/\beta_1 &-\gamma_1^2/\beta_1^2
\\ \beta_2/\gamma_2 & 1 &\gamma_2/\beta_2 &-\gamma_2^2/\beta_2^2
\\ \beta_3/\gamma_3 & 1 &\gamma_3/\beta_3 &-\gamma_3^2/\beta_3^2
\\ \beta_4/\gamma_4 & 1 &\gamma_4/\beta_4 &-\gamma_4^2/\beta_4^2
\end{array}\right|
\neq 0 .
\label{cond3}
\end{equation}
Then $t,r,x,z$ can be expressed through
$\Phi_1,\Phi_2,\Phi_3,\Phi_4$ and the same for
$\Phi_5,\ldots,\Phi_n$, so that $\Phi_j$ for $j=1,2,3,4$ can be
chosen as new independent variables in (\ref{1stint2}) and the
equation (\ref{1stint2}) takes the form (\ref{form2nd}).

In our case solutions of (\ref{legheav2}) are already determined
in (\ref{sol3}) up to arbitrary constants by solving second-order
linear equations together with the Legendre-transformed second
heavenly equation and hence, having no functional arbitrariness,
they cannot satisfy in addition the first-order equation
(\ref{1stint2}). Thus for any choice of functions
$q,c,e,\rho,\sigma,\kappa$, depending on $z,w=-u_t$, arbitrary
constants $a,k$ and the potential $P(x,r,z,t)$ satisfying
(\ref{potP}) the equation (\ref{1stint2}) is not an identity but
an equation of the form (\ref{form2nd}). This implies a dependence
of the independent variables which is a contradiction that proves
nonexistence of the Killing vectors for the metric
(\ref{legmetr2}) where the potential $u$ is determined by
(\ref{sol3}) with $n\ge 4$ and the conditions (\ref{cond2}) or
(\ref{cond3}) are satisfied.

If $n\le 3$ the above reasoning obviously does not work and Killing vectors
may exist.

Thus we have proved the following theorem.
\begin{theorem}
Metric (\ref{legmetr2}) with $u$ defined by (\ref{sol3})
with the conditions (\ref{alp_del}) or (\ref{alp_del2}),
$n\ge 4$, $\beta_1\beta_2\beta_3\beta_4 \neq 0$ and satisfying the
non-degeneracy conditions (\ref{cond2}) or (\ref{cond3}) respectively
admits no Killing vectors.
\end{theorem}

\section{Conclusions and discussion}

We have shown that our method of partner symmetries, worked out
initially for the elliptic complex Monge-Amp\`ere equation, can be
extended to the hyperbolic complex Monge-Amp\`ere equation and the
second heavenly equation of Pleba\~nski. This method enables us to
construct second order differential constraints which select
certain particular sets of solutions invariant with respect to
non-local symmetries and hence non-invariant in the usual sense.
The advantage of the method is that we deal only with local point
symmetries, which build up non-local symmetries, and do not need
to work with non-local symmetries explicitly. We found some simple
choices of these local partner symmetries for which the Legendre
transformation converts the original heavenly equation together
with differential constraints to linear equations. We have found
their generically non-invariant solutions, dependent on all four
variables, and hence new classes of 4-dimensional heavenly metrics
without Killing vectors.

The idea of obtaining non-invariant solutions as invariant
solutions with respect to non-local symmetry for the hierarchy
associated to the second heavenly equation was also suggested by
Dunajski and Mason. However, their `hidden symmetries' constitute
a very special class of symmetries which can be generated from
local symmetries by repeated applications of the recursion
operator and hence have a characteristic property that they can be
mapped back to a certain local symmetry by some power of the
inverse recursion operator. As a consequence they have at least
six differential second-order constraints implied by the hidden
symmetries. A class of non-local symmetries constructed from
partner symmetries is much more extensive because it consists of
symmetries which are linear combinations of local symmetries and
those generated from local symmetries by the recursion operators,
so that they cannot be mapped to a local symmetry by the action of
recursion operators. This additional freedom in symmetries results
in wider classes of solutions of the heavenly equations since the
number of additional differential constraints implied by partner
symmetries is typically three, which is less than six in the case
of hidden symmetries.

The crucial point of our method is the possibility of linearizing
the field equation together with constraints by the Legendre
transformation. We have found that this is possible for some
simple choices of partner symmetries. An important problem, with
which we are occupied, is to work out a criterion which would
choose such symmetries for which the linearization is possible.

Another project, that we are working on, is to construct general
classes of equations for which our method of partner symmetries
could be applied. Characteristic features of such equations should
include the divergence structure of the determining equation for
symmetries and the condition that a potential for any symmetry
should again be a symmetry. We plan to return to this problem soon
in a future publication.

\section{Acknowledgements}

One of us, MBS, thanks Thomas Wolf for providing me the latest
versions of his Reduce packages Liepde and Applysym for
calculating symmetry algebras and performing Legendre
transformations. We thank the referees for their criticism
and profound remarks that, as we hope, served to a substantial improvement
of our paper.


\begin{thebibliography}{9}
\bibitem{pleb}
Pleba\~nski J F 1975 {\it J. Math. Phys.} {\bf 16} 2395--2402
\bibitem{ahs}
Atiyah M F, Hitchin N J and Singer I M 1978  {\it Proc. Roy. Soc.
A} {\bf 362} 452
\bibitem{mns}
Malykh A A, Nutku Y and Sheftel M B
2003 {\it J. Phys. A:  Math. Gen.} {\bf 36} 10023--10037
\bibitem{mnsgr}
Malykh A A, Nutku Y and Sheftel M B 2003 {\it Class. Quantum
Grav.} {\bf 20} L263--L266
\bibitem{dm}
Dunajski M and Mason L J 2003 {\it J. Math. Phys.} {\bf 44}
3430--3454; arXiv:math.DG/0301171 v2 16 Jun 2003.
\bibitem{ivroc}
Ivanov I T and Ro\u cek M 1996 {\it Comm. Math. Phys.} {\bf 182}
291--302
\bibitem{gol} Goldblatt E 1994 {\it Gen. Rel. and Grav.} {\bf 26} 979
\bibitem{an} Aliev A N and Nutku Y 1999 {\it Class. Quantum Gravity} {\bf 16}
189
\bibitem{lp}
Lechtenfeld O and Popov A D 2000 {\it Int. J. Mod. Phys. A} {\bf
15} 4191--4236
\bibitem{bw}
Boyer C P and Winternitz P 1989 {\it J. Math. Phys.}  {\bf 30}
1081--1094
\bibitem{olv}
Olver P 1986 {\it Applications of Lie Groups to Differential
Equations} (New York: Springer)
\bibitem{masnew}
Mason L J and Newman E T 1989 {\it Commun. Math. Phys.} {\bf 121}
659--668
\bibitem{mw}
Mason L J and Woodhouse N M J 1996 {\it Integrability,
self-duality, and twistor theory} (Oxford: Clarendon Press)
\bibitem{dm2000}
Dunajski M and Mason L J 2000 {\it Commun. Math. Phys.} {\bf 213}
641--672
\bibitem{LW}
Levi D and Winternitz P 1989 {\it J. Phys. A: Math. Gen.} {\bf 22} 2915--2924
\end{thebibliography}
\end{document}